\documentclass[twocolumn,letter]{jpsj2}

\def\runtitle{Neutron Diffraction Study of the Pressure-Induced Magnetic Ordering in TlCuCl$_3$}
\def\runauthor{Akira \textsc{Oosawa} {\it et al.}}

\title{Neutron Diffraction Study of the Pressure-Induced Magnetic Ordering in the Spin Gap System TlCuCl$_3$}

\author{Akira \textsc{Oosawa}\thanks{E-mail address: a-oosawa@neutrons.tokai.jaeri.go.jp.}, Masashi \textsc{Fujisawa}$^{1}$, Toyotaka \textsc{Osakabe}, Kazuhisa \textsc{Kakurai} and Hidekazu \textsc{Tanaka}$^2$}

\inst{Advanced Science Research Center, Japan Atomic Energy Research Institute, Tokai, Ibaraki 319-1195, Japan\\
$^1$Department of Physics, Tokyo Institute of Technology, Oh-okayama, Meguro-ku, Tokyo 152-8551, Japan\\
$^2$Research Center for Low Temperature Physics, Tokyo Institute of Technology, Oh-okayama, Meguro-ku, Tokyo 152-8551, Japan}

\recdate{\today}

\abst{Neutron elastic scattering measurements have been performed under the hydrostatic pressure in order to investigate the spin structure of the pressure-induced magnetic ordering in the spin gap system TlCuCl$_3$. Below the ordering temperature $T_{\rm N}=16.9$ K for the hydrostatic pressure $P=1.48$ GPa, magnetic Bragg reflections were observed at the reciprocal lattice points ${\mib Q}=(h, 0, l)$ with integer $h$ and odd $l$, which are equivalent to those points with the lowest magnetic excitation energy at ambient pressure. This indicates that the spin gap closes due to the applied pressure. The spin structure of the pressure-induced magnetic ordered state for $P=1.48$ GPa was determined.}

\kword{TlCuCl$_3$, spin gap, pressure-induced magnetic ordering, spin structure, neutron elastic scattering}

\begin{document}
\sloppy
\maketitle

The spin gap system is a magnetic system having the singlet spin liquid ground state with a finite excitation gap \cite{Dagotto}. Recently, the magnetic ordering induced by modifying the gapped ground state by an external field or impurity doping has been energetically investigated. When a magnetic field which is higher than the gap field $H_{\rm g}$ corresponding to the energy gap $\Delta = g \mu_{\rm B} H_{\rm g}$ is applied in a spin gap system, the energy gap vanishes, and the system can undergo the magnetic ordering with the help of the three-dimensional (3D) interactions. On the other hand, when nonmagnetic ions are substituted for magnetic ions in a spin gap system, the singlet ground state is disturbed, so that staggered moments are induced around the impurities. If the induced moments interact through effective exchange interactions, which are mediated by intermediate singlet spins, the 3D long-range order can arise. Such field-induced and impurity-induced magnetic orderings were observed in many spin gap systems \cite{Hammer,Honda,Manaka,Masuda1,Azuma,Uchiyama}. \par
The application of pressure is another method to control the quantum magnetism including the spin gap and the spin-Peierls transition. Some remarkable pressure effects have been observed in some quantum spin systems. The pressure-induced magnetic ordering were observed in Cu$_2$(C$_5$H$_{12}$N$_2$)$_2$Cl$_4$ \cite{Mito}, which was first assumed to be an $S=1/2$ Heisenberg antiferromagnetic two-leg ladder \cite{Chaboussant} and was later characterized as an frustrated 3D spin gap system \cite{Stone}. In Cu$_2$(C$_5$H$_{12}$N$_2$)$_2$Cl$_4$, a part of singlet spin pairs are broken by the applied pressure to become paramagnetic spins, and the magnetic ordering similar to the impurity-induced antiferromagnetic ordering occurs in the rest of spins \cite{Mito}. The spin gap remains even in the ordered state. \par
In a well-known inorganic spin-Peierls (SP) material CuGeO$_3$, the SP phase was enhanced by applied pressure \cite{Takahashi}, so that the revival of the SP phase was observed for highly-Mg-doped CuGeO$_3$, in which the SP transition does not occur and the impurity-induced antiferromagnetic ordering only occurs at the low temperature for ambient pressure \cite{Masuda2}. \par
This paper is concerned with the pressure-induced magnetic ordering in the spin gap system TlCuCl$_3$. \par
We summarize the physical properties of TlCuCl$_3$ at ambient pressure obtained up to date. This system has the monoclinic structure (space group $P2_1/c$) \cite{Takatsu}. The crystal structure consists of planar dimers of Cu$_2$Cl$_6$. The dimers form infinite double chain along the crystallographic $a$-axis. These chains are located at the corners and the center of the unit cell in the $b-c$ plane, and are separated by Tl$^+$ ions. The magnetic ground state of TlCuCl$_3$ is the spin singlet \cite{Takatsu} with the excitation gap $\Delta=7.7$ K \cite{Shiramura,Oosawamag}. The origin of the spin gap is the strong antiferromagnetic spin dimer in the chemical dimer Cu$_2$Cl$_6$, and the neighboring spin dimers couple via strong 3D interdimer interactions along the double chain and in the $(1, 0, -2)$ plane, in which the hole orbitals of Cu$^{2+}$ spread \cite{Oosawainela,Cavadini}. \par
The field-induced magnetic ordering was observed in TlCuCl$_3$ \cite{Oosawamag,Oosawaheat,Tanakaela}. For the magnetic ordering, two remarkable features have been found. One is that the magnetization has the cusplike minimum at the transition temperature $T_{\rm N}$ for $H > H_{\rm g}$. The other is that the phase boundary between the paramagnetic phase and the ordered phase can be described by the power law. These features cannot be described by the mean-field approach from the real space \cite{Tachiki1,Tachiki2}. Nikuni {\it et al.} \cite{Nikuni} demonstrated that the field-induced magnetic ordering in TlCuCl$_{3}$ can be represented as a Bose-Einstein condensation (BEC) of spin triplets (magnons), and that the above-mentioned two features are qualitatively well described by the magnon BEC theory based on the Hartree-Fock (HF) approximation. \par
\begin{figure}[t]
\begin{center}
\includegraphics[width=80mm]{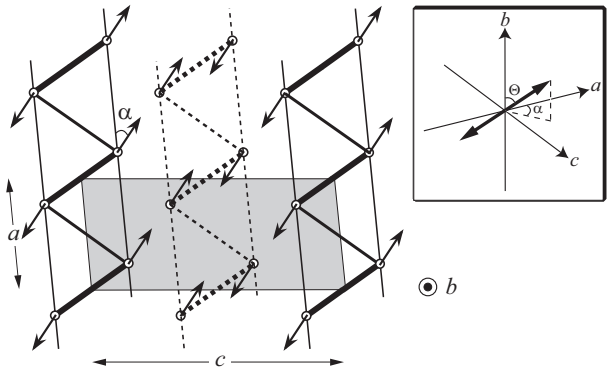}%
\end{center}
\caption{Projection of the spin structure observed in the field-induced magnetic ordered phase of TlCuCl$_3$ for ${\mib H} \parallel b$, the impurity-induced antiferromagnetic ordered phase of Tl(Cu$_{0.97}$Mg$_{0.03}$)Cl$_3$ and the pressure-induced magnetic ordered phase for $P=1.48$ GPa in TlCuCl$_3$ onto the $a-c$ plane. The double chain located at the corner and the center of the chemical unit cell in the $b-c$ plane are represented by solid and dashed lines, respectively. The shaded area is the chemical unit cell in the $a-c$ plane. The inset shows the inclination of the ordered spin toward the $b$-axis. The angle $\alpha$ denotes the angle between the $a$-axis and the spin component projected onto the $a-c$ plane. The angle  $\Theta$ is the angle between the spin and the $b$-axis. \label{spinstructure}}
\end{figure}

If the magnons undergo BEC at an ordering vector ${\mib Q}_0$ for $H>H_{\rm g}$, then the transverse spin components have long-range order, which is characterized by the same wave vector ${\mib Q}_0$. The transverse magnetic ordering predicted by the theory \cite{Tachiki1,Tachiki2,Nikuni} was confirmed by the neutron elastic scattering experiments in TlCuCl$_3$ \cite{Tanakaela}. The spin structure obtained for ${\mib H} \parallel b$ is shown in Fig. \ref{spinstructure} with $\Theta=90^{\circ}$ and $\alpha=39^{\circ}$. The spins lie in the $a-c$ plane which is perpendicular to the applied field. Spins on the same dimers represented by thick lines in Fig. \ref{spinstructure} are antiparallel. Spins are arranged in parallel along a leg in the double chain, and make an angle of $\alpha$ with the $a$-axis. The spins on the same legs in the double chains located at the corner and the center of the unit cell in the $b-c$ plane are antiparallel. \par
The impurity-induced antiferromagnetic ordering was also observed by the magnetization measurements in Tl(Cu$_{1-x}$Mg$_x$)Cl$_3$ \cite{OosawaMgsus}. By means of neutron scattering measurements, the ordering was confirmed in Tl(Cu$_{0.97}$Mg$_{0.03}$)Cl$_3$ below $T_{\rm N}=3.45$ K, and the spin structure of the ordered phase was determined \cite{OosawaMgneu}. The spin structure on average is equivalent to that shown in Fig. \ref{spinstructure} with $\Theta=90^{\circ}$ and $\alpha=34.0^{\circ} \pm 4.7^{\circ}$. The effective magnetic moment averaged per site at $T=1.4$ K was evaluated as $\langle m_{\rm eff} \rangle =g{\mu}_{\rm B}\langle S_{\rm eff} \rangle = 0.12(1)$ ${\mu}_{\rm B}$. \par
Quite recently, the magnetization measurements were performed under the hydrostatic pressure up to $P\approx 0.8$ GPa in TlCuCl$_3$, and the antiferromagnetic ordering was observed at zero field for $P > 0.4$ GPa \cite{TanakaLT}, {\it e.g.}, the ordering temperature $T_{\rm N}$ at $P\approx 0.8$ GPa is $T_{\rm N} \approx 11$ K. It was found that the easy axis of the magnetic moments is close to the [2, 0, 1] direction, which is parallel to both $(0, 1, 0)$ and $(1, 0, -2)$ planes, and makes an angle of 51$^{\circ}$ with the $a$-axis. Thus, TlCuCl$_3$ undergoes phase transitions induced by magnetic field, impurities and hydrostatic pressure. In order to investigate the spin structure in the pressure-induced magnetic ordered phase and the mechanism leading to the ordering, we performed neutron elastic scattering experiments under the high hydrostatic pressure. \par
The preparation of the single crystal of TlCuCl$_3$ has been reported in reference \citen{Oosawamag}. The sample with 0.2 cm$^3$ was set in the McWhan type high pressure cell (HPCNS-MCW$^{\circledR}$, Oval Co., Ltd.) \cite{McWhan}. As the pressure transmitting medium, a mixture of Fluorinert FC70 and FC77 was used. The applied hydrostatic pressure of $P=1.48$ GPa at low temperature was determined from the pressure dependence of the lattice constants of a NaCl crystal in the sample space. Neutron elastic scattering measurements were performed using the JAERI-TAS1 installed at JRR-3M, in Tokai. The constant-${\mib k}_i$ mode was taken with a fixed incident neutron energy $E_i$ of 14.7 meV. Because the size of the sample has to be small due to the small sample space, collimations were set as open-80'-80'-80' in order to gain intensity. Sapphire and pyrolytic graphite filters were placed to suppress the background by high energy neutrons, and higher order contaminations, respectively. The sample was mounted in the cryostat with its $a^*$- and $c^*$-axes in the scattering plane. The crystallographic parameters were determined as $a^*=1.6402$ 1/$\rm{\AA}$, $c^*=0.72843$ 1/$\rm{\AA}$ and $\cos\beta^*=0.0861$ at helium temperatures and at $P=1.48$ GPa. The lattice constants $a$ and $c$ become shorter by 1.5 \% and the angle $\beta$ becomes closer to 90$^{\circ}$ as compared with those at ambient pressure \cite{Oosawainela}. \par
\begin{figure}[t]
\begin{center}
\includegraphics[width=80mm]{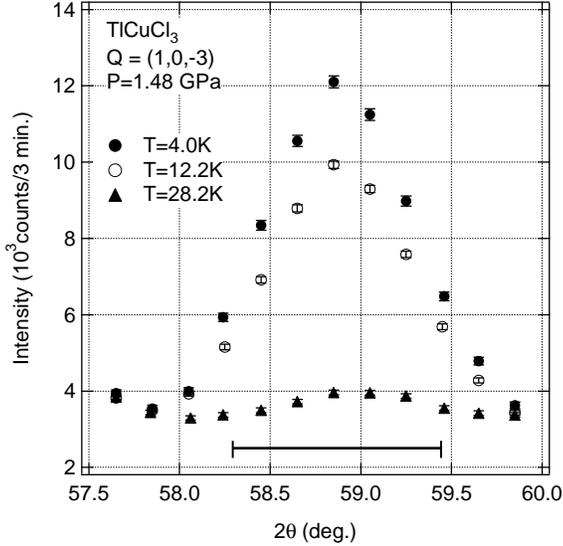}%
\end{center}
\caption{${\theta}-2{\theta}$ scans for ${\mib Q} = (1, 0, -3)$ reflection measured at $T=4.0$, 12.2 and 28.2 K at $P=1.48$ GPa in TlCuCl$_3$. The horizontal bar denotes the calculated instrumental resolution width. \label{Q10-3profile}}
\end{figure}
\begin{figure}[t]
\begin{center}
\includegraphics[width=80mm]{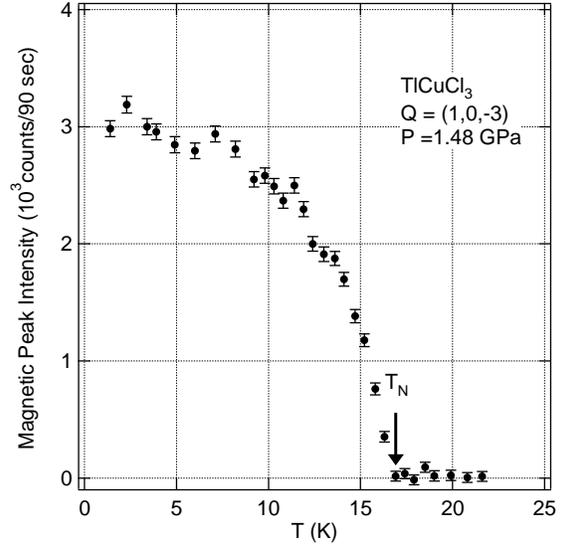}%
\end{center}
\caption{Temperature dependence of the magnetic Bragg peak intensity for ${\mib Q}=(1, 0, -3)$ reflection measured at $P=1.48$ GPa in TlCuCl$_3$. \label{Q10-3temdep}}
\end{figure}

Figure \ref{Q10-3profile} shows the ${\theta}-2{\theta}$ scans for ${\mib Q} = (1, 0, -3)$ reflection measured at $T=4.0$, 12.2 and 28.2 K for $P=1.48$ GPa in TlCuCl$_3$. With decreasing temperature, the increase of the intensity of the reflection with resolution limited width can be clearly seen. Figure \ref{Q10-3temdep} shows the temperature dependence of the Bragg peak intensity at ${\mib Q}=(1, 0, -3)$ reflection for $P=1.48$ GPa in TlCuCl$_3$. Since this measurement was carried out on a condition different from that for the previous ${\theta}-2{\theta}$ scans, the neutron counts for magnetic Bragg reflection are not equal to those shown in Fig. \ref{Q10-3profile}. However, we confirmed that the observed magnetic Bragg intensities are consistent with one another by comparing the nuclear Bragg intensities observed in both measurements. The rapid increase of the intensity with no pronounced diffuse scattering is observed below $T_{\rm N}=16.9$ K. Because it has been already observed that the pressure-induced magnetic ordering occurs at $T_{\rm N} \approx 11$ K for $P\approx 0.8$ GPa in TlCuCl$_3$ \cite{TanakaLT}, we can conclude that the ${\mib Q}=(1, 0, -3)$ reflection is the magnetic Bragg reflection indicative of the pressure-induced magnetic ordering. The magnetic Bragg reflections were observed at ${\mib Q}=(h, 0, l)$ with integer $h$ and odd $l$. These reciprocal points are equivalent to those with the lowest magnetic excitation energy at ambient pressure \cite{Oosawainela,Cavadini}. Ferromagnetic Bragg reflections could not be detected in the present measurements. These results indicates that the spin gap corresponding to the lowest magnetic excitation energy closes under the pressure. Thus, it can be deduced that the interdimer interactions are relatively enhanced against the intradimer interaction due to the applied pressure, so that the spin gap is reduced and closed completely. The mechanism leading to the pressure-induced magnetic ordering in TlCuCl$_3$ is different from that in Cu$_2$(C$_5$H$_{12}$N$_2$)$_2$Cl$_4$ \cite{Mito}. These reciprocal points are also equivalent to those for the magnetic Bragg peaks indicative of the field-induced magnetic ordering for ${\mib H} \parallel b$ in TlCuCl$_3$ \cite{Tanakaela} and the impurity-induced antiferromagnetic ordering in Tl(Cu$_{0.97}$Mg$_{0.03}$)Cl$_3$ \cite{OosawaMgneu}.\par
In order to determine the spin structure of the pressure-induced magnetic ordered phase in TlCuCl$_3$, we measured the integrated intensities of nine Bragg reflections at $T=4.0$, 12.2 and 28.2 K. The integrated intensities were obtained by the ${\theta}-2{\theta}$ scans. The results are summarized in Table \ref{table1} together with the calculated intensities. The intensities of nuclear Bragg reflections were also measured to estimate the magnitude of the magnetic moment per site. The nuclear peaks were observed at ${\mib Q}=(h, 0, l)$ with even $l$, as expected from the space group $P2_1/c$. However, very weak nuclear peaks were also observed for odd $l$ as observed at ambient pressure \cite{Tanakaela}. To refine the magnetic structure, we used the atomic coordinates of TlCuCl$_3$ at ambient pressure \cite{Tanakaela} and the nuclear scattering lengths $b_{\rm Tl}=0.878$, $b_{\rm Cu}=0.772$ and $b_{\rm Cl}=0.958$ with the unit of 10$^{-12}$ cm \cite{Sears}. The magnetic form factors of Cu$^{2+}$ were taken from reference \citen{Brown}. The extinction effect was evaluated by comparing observed and calculated intensities for various nuclear Bragg reflections. \par
As a result, the spin structure as shown in Fig. \ref{spinstructure} with $\alpha=42.6^{\circ} \pm 1.4^{\circ}$ and $\Theta=90.0^{\circ} \pm 9.0^{\circ}$ for $T=12.2$ K, which is almost same as those obtained in the field-induced and impurity-induced magnetic ordered phases, and that with $\alpha=49.5^{\circ} \pm 2.4^{\circ}$ and $\Theta=58.0^{\circ} \pm 3.4^{\circ}$ for $T=4.0$ K were obtained, respectively. This result means that the ordered moments lie in the $a-c$ plane just below $T_{\rm N}$, and that they incline toward the $b$-axis with decreasing temperature. However, in the temperature dependence of the magnetic Bragg peak intensity shown in Fig. \ref{Q10-3temdep}, there is no apparent anomaly indicative of an additional phase transition, at which the ordered moments start to incline toward the $b$-axis. According to the magnetization measurements up to $P\approx 0.8$ GPa, the easy-axis is close to the [2, 0, 1] direction down to $T=1.8$ K \cite{TanakaLT}. Thus, it is considered that the inclination of ordered moments arises for $P > 1$ GPa. \par
Within the present measurements, it is not clear whether the inclination of ordered moments occurs gradually below $T_{\rm N}$, or as a phase transition. The polarized neutron scattering technique will be useful to settle the question, because the spin components parallel and perpendicular to the scattering plane ($a-c$ plane) can be observed independently \cite{Moon}. This experiment is planned for near future. \par
Comparing magnetic peak intensities with those of nuclear reflections, the magnitude of the ordered moment was evaluated as $\langle m \rangle =g{\mu}_{\rm B}\langle S \rangle = 0.64(4)$ ${\mu}_{\rm B}$ at $T=4.0$ K and $\langle m \rangle = 0.51(3)$ ${\mu}_{\rm B}$ at $T=12.2$ K. \par
The magnitude of the ordered moment at $T=4.0$ K is 64(4) \% of the full moment of Cu$^{2+}$ and is much larger than those observed in the field-induced magnetic ordering for ${\mib H} \parallel b$ in TlCuCl$_3$ \cite{Tanakaela} and the impurity-induced antiferromagnetic ordering in Tl(Cu$_{0.97}$Mg$_{0.03}$)Cl$_3$ \cite{OosawaMgneu}. This indicates that the spin gap is completely collapsed at $P=1.48$ GPa and the triplet states of the spin dimer make a large contribution to the ground state. \par
\begin{fulltable}[t]
\caption{Observed and calculated magnetic Bragg peak intensities at $T=12.2$ and 4.0 K for $P=1.48$ GPa in TlCuCl$_3$. The intensities are normalized to the (0, 0, 1)$_{\rm M}$ reflection. $R$ is the reliability factor given by $R=\sum_{h,k,l}|I_{\rm cal}-I_{\rm obs}|/\sum_{h,k,l}I_{\rm obs}$.\label{table1}}
\begin{center}
\begin{fulltabular}{ccccccccc} \hline
& \hspace{1cm} & \multicolumn{3}{c}{\hspace{0.5cm}$T=12.2$ K} & \hspace{3cm} & \multicolumn{3}{c}{\hspace{0.5cm}$T=4.0$ K}\\ \hline
$(h, k, l)$ & \hspace{1cm} & $I_{\rm obs}$ & \hspace{0.8cm} & $I_{\rm cal}$ & \hspace{2cm} & $I_{\rm obs}$ & \hspace{0.8cm} & $I_{\rm cal}$ \\ \hline
(0, 0, 1)$_{\rm M}$ & \hspace{1cm} & 1 $\pm$ 0.051 & \hspace{0.8cm} & 1 & \hspace{3cm} & 1 $\pm$ 0.051 & \hspace{0.8cm} & 1\\
(0, 0, 3)$_{\rm M}$ & \hspace{1cm} & 0.018 $\pm$ 0.004 & \hspace{0.8cm} & 0.014 & \hspace{3cm} & 0.007 $\pm$ 0.003 & \hspace{0.8cm} & 0.014\\
(0, 0, 5)$_{\rm M}$ & \hspace{1cm} & 0.085 $\pm$ 0.007 & \hspace{0.8cm} & 0.103 & \hspace{3cm} & 0.075 $\pm$ 0.006 & \hspace{0.8cm} & 0.103\\
(1, 0, 1)$_{\rm M}$ & \hspace{1cm} & 0.014 $\pm$ 0.006 & \hspace{0.8cm} & 0.021 & \hspace{3cm} & 0.104 $\pm$ 0.009 & \hspace{0.8cm} & 0.113\\
(1, 0, $-1$)$_{\rm M}$ & \hspace{1cm} & 0.127 $\pm$ 0.014 & \hspace{0.8cm} & 0.153 & \hspace{3cm} & 0.135 $\pm$ 0.016 & \hspace{0.8cm} & 0.165\\
(1, 0, 3)$_{\rm M}$ & \hspace{1cm} & 0.037 $\pm$ 0.009 & \hspace{0.8cm} & 0.014 & \hspace{3cm} & 0.083 $\pm$ 0.004 & \hspace{0.8cm} & 0.085\\
(1, 0, $-3$)$_{\rm M}$ & \hspace{1cm} & 0.448 $\pm$ 0.009 & \hspace{0.8cm} & 0.413 & \hspace{3cm} & 0.371 $\pm$ 0.007 & \hspace{0.8cm} & 0.375\\
(2, 0, 1)$_{\rm M}$ & \hspace{1cm} & 0.006 $\pm$ 0.003 & \hspace{0.8cm} & 0.021 & \hspace{3cm} & 0.057 $\pm$ 0.005 & \hspace{0.8cm} & 0.051\\
(2, 0, $-1$)$_{\rm M}$ & \hspace{1cm} & 0.114 $\pm$ 0.007 & \hspace{0.8cm} & 0.127 & \hspace{3cm} & 0.191 $\pm$ 0.008 & \hspace{0.8cm} & 0.157\\ \hline
$R$ & \hspace{1cm} & & \hspace{0.8cm} & 0.08 & \hspace{3cm} & & \hspace{0.8cm} & 0.06\\ \hline
\end{fulltabular}
\end{center}
\end{fulltable}
In conclusion, we have presented the first neutron elastic scattering results on the spin gap system TlCuCl$_3$ under the hydrostatic pressure. Below $T_{\rm N}=16.9$ K for $P=1.48$ GPa, the magnetic Bragg reflections indicative of the magnetic ordering were observed at the reciprocal lattice points ${\mib Q}=(h, 0, l)$ with integer $h$ and odd $l$, which are equivalent to those points with the lowest magnetic excitation energy at ambient pressure. This indicates that the magnetic ordering observed arises from the closing of the spin gap under the pressure. These reciprocal points are also equivalent to those for the magnetic Bragg peaks indicative of the field-induced magnetic ordering for ${\mib H} \parallel b$ in TlCuCl$_3$ and the impurity-induced antiferromagnetic ordering in Tl(Cu$_{0.97}$Mg$_{0.03}$)Cl$_3$. The spin structure of the pressure-induced ordered phase in TlCuCl$_3$ for $P=1.48$ GPa was determined as shown in Fig. \ref{spinstructure} with $\alpha=42.6^{\circ} \pm 1.4^{\circ}$ and $\Theta=90.0^{\circ} \pm 9.0^{\circ}$ for $T=12.2$ K and $\alpha=49.5^{\circ} \pm 2.4^{\circ}$ and $\Theta=58.0^{\circ} \pm 3.4^{\circ}$ for $T=4.0$ K. This implies that the ordered moments lying originally in the $a-c$ plane just below $T_{\rm N}$ incline toward the $b$-axis at lower temperatures. The magnitude of the ordered moment was determined as $\langle m \rangle =0.64(4)$ $\mu_{\rm B}$ at $T=4.0$ K. \par
We acknowledge Y. Shimojo for his technical support. This work was supported by the Toray Science Foundation and a Grant-in-Aid for Scientific Research on Priority Areas (B) from the Ministry of Education, Culture, Sports, Science and Technology of Japan. \par

\end{document}